\documentclass[twocolumn,superscriptaddress,showpacs,amsfonts,amsmath,amssymb]{revtex4-1}
\pdfoutput=1

\usepackage{graphicx}
\usepackage{amsmath}
\usepackage{amsfonts}
\usepackage{amssymb}
\usepackage{epsfig}
\usepackage{color}
\usepackage{dsfont}

\newcommand{\abs}[1]{\left\vert#1\right\vert}

\newcommand{\bra}[1]{\left\langle#1\right\vert}
\newcommand{\ket}[1]{\left\vert#1\right\rangle}
\newcommand{\Tr}[1]{\text{Tr}\left\{#1\right\}}

\newcommand\braket[2]{\left.\left\langle#1\right|#2\right\rangle}

\begin{document}

\author{G.~Francica}
\affiliation{CNR-SPIN, I-84084 Fisciano (Salerno), Italy}

\author{P.~Gentile}
\affiliation{CNR-SPIN, I-84084 Fisciano (Salerno), Italy}\affiliation{Dipartimento di Fisica ``E. R. Caianiello'', Universit\`a di Salerno, I-84084 Fisciano (Salerno), Italy}

\author{M.~Cuoco}
\affiliation{CNR-SPIN, I-84084 Fisciano (Salerno), Italy}\affiliation{Dipartimento di Fisica ``E. R. Caianiello'', Universit\`a di Salerno, I-84084 Fisciano (Salerno), Italy}

\title{Spin-Orbital Textures and Interferometers by 
Nanoscale \\ Geometric Design of Quantum Rings with Orbital Rashba Coupling}
\begin{abstract}
We derive an effective continuum model for describing the propagation of electrons in ballistic one-dimensional curved nanostructure which are marked by a strong interplay of spin-orbital degrees of freedom due to local electronic states with $d-$orbital symmetry, atomic spin-orbit and orbital Rashba couplings.
We demonstrate how a planar inhomogeneous spatial curvature of the nanochannel can generate both spin and orbital textures represented through loops on the corresponding Bloch spheres. In particular, we employ the paradigmatic case of an elliptically shaped quantum ring to investigate the role of geometry in steering the electron transport at low energy. We find that in this regime spin and orbital textures exhibit equal windings around the radial or out-of-plane directions. Remarkably, the spin and orbital windings can be not only tuned through a modification of the nanoscale shape, as for the spin Rashba effect, but also via a change in the electron density by electrical gating. We show that the presence of textures is related to patterns in the geometric Aharonov-Anandan phase acquired in a cycle within the quantum ring. Furthermore, differently from the single band Rashba-coupled nanoring, a manipulation of the conductance can be achieved by electron density variation and, indirectly, through changes of the strength of the inversion asymmetric interactions.
\end{abstract}

\maketitle

\section{Introduction}

The manipulation and control of the electron spin set fundamental challenges for the development of emergent technologies and quantum engineering. These mainly include spin-based quantum computation and encryption systems, spintronics devices, and everything that can employ quantum degrees of freedom beyond the electron charge to achieve exceptional performances in speed, energy requirements, and functionality. 
While the most direct control over the spins has been exploiting the magnetic field, novel paradigms are more and more common with significant step forward in next generation of magnetic-field free spin-based technology. Along this direction, the spin-orbit coupling, as for instance the Rashba interaction ~\cite{Rashba1960} in low-dimensional systems with structure inversion asymmetry, represents a key microscopic ingredient for an all-electrical intrinsic control over the spin through the electron propagation, and at the same time as a source of spin control of the electron trajectories ~\cite{manchon15}. 

On a general ground, there are two key interactions that mark the inversion asymmetric microscopic environments, i.e. the Rashba spin-orbit coupling occurring at the 
surface or interface of heterostructures, and the Dresselhaus interaction arising from the inversion asymmetry in the bulk of the host material~\cite{dresselhaus}. 
Apart from these fundamental couplings, it has been recently realized that spin-momentum locking can also occur as a consequence of the inversion symmetry breaking arising from the orbital polarization of electrons in solids which is, then, linked with the spin-sector by the atomic spin-orbit coupling. 
The role of spin and orbital polarization in materials can lead to different scenarios of the manifestation of inversion symmetry breaking when compared to the canonical spin-Rashba effect, leading to the so-called orbital-driven Rashba coupling~\cite{park11}. The orbital Rashba (OR) effect can yield two-dimensional chiral orbital textures and orbital dependent spin-vector via the spin-orbit coupling~\cite{park11,park13,kim14,hong15,kim12,park12,park12a,fukaya2}.
Evidences of anomalous energy splitting have been found in a large variety of surfaces, i.e.\ Au(111), Pb/Ag(111)~\cite{el-kareh14}, Bi/Ag(111)~\cite{schirone15}, etc.\ as well as in transition metal oxides based interfaces, i.e.\ LaAlO$_3$-SrTiO$_3$~\cite{King2014,Nakamura2012}.
Such phenomenology is typically encountered in multi-orbital materials where the combination of the atomic spin-orbit interaction with orbitally driven inversion symmetry-breaking sources effectively generates emergent asymmetric spin-orbital interactions within the electronic manifold close to the Fermi level. 

Orbital degrees of freedom are crucial microscopic ingredients in quantum 
materials when considering the degeneracy of $d$-bands of 
transition elements not being completely removed by the crystal 
distortions or due to the intrinsic spin-orbital entanglement~\cite{ole12} triggered by the atomic spin-orbit coupling. 
Motivated by the search for innovative paths to successfully achieve an electron spin control, here we investigate the interplay of the orbital degree of freedom and the geometry of the quantum system in which electrons reside with applications based on quantum rings that can be potentially set out at the interface of oxide band insulators. 
In this context, LaAlO$_3$-SrTiO$_3$ \cite{ohmoto} is a paradigmatic example of orbitally controlled quantum material because the filling of titanium $d-$ orbitals can be electrically tuned leading to an orbital dependent \cite{salluzzo} two-dimensional electron gas (2DEG) with high-mobility. Furthermore, the combination of atomic spin-orbit at the transition element and the gate tunable and orbital dependent inversion asymmetric interaction leads to a highly non-trivial spin-orbital splitting in the reciprocal space.
The OR interaction indeed manifests through mixing of orbitals on neighboring atoms, i.e. $d_{xy}$ with $(d_{xz}$,$d_{yz})$, that would not overlap in an inversion symmetric configuration. 
Remarkably, electric field control of 2DEG transition metal-oxide-based structures have recently enabled the exploration of nanoscale electron quantum transport \cite{caviglia18,bergeal} thus emphasizing the prospect of advanced quantum engineering including also the possibility of topological superconducting phases \cite{schmalian,loder,fukaya1,fukaya2,perroni}. 

For the canonical Rashba coupling the electron spin can be successfully guided by non-trivial geometric curvature of the semiconducting nanochannel. In these physical systems a manipulation of the electronic states, through the corresponding spin geometric phase, has been experimentally demonstrated \cite{nagasawa1,nagasawa2}
with remarkable perspectives of achieving topological spin engineering \cite{saarikoski,reynoso}.
The potential of the union of inversion symmetry breaking and nanoscale shaping indeed yields augmenting paths for topological states \cite{saarikoski,reynoso,gentile,ying} and spin-transport \cite{nagasawa2, frustaglia1, koga, bercioux, koga1, Qu}. There, the Rashba spin-orbit driving fields act as spatially inhomogeneous geometrical torque controlling both the spin-orientation and its spin-phase through non-trivial spin windings \cite{saarikoski,reynoso,ying}.

Taking into account such a geometrical driving of the electron spin through the conventional spin Rashba coupling, here we ask whether the conspiracy of the orbital Rashba interaction and the atomic spin-orbit coupling in nanoscale shaped quantum rings \cite{fomin} can lead to unconventional spin-orbital textures and innovative control mechanisms for the quantum transport.
Remarkably, we find that the geometric curvature of a shape deformed one-dimensional nanostructure can both steer the electron spin and the orbital angular momentum and, in turn, the quantum geometric phase that an electron acquires when moving in a closed circuit. These results provide a non standard view on the way to manipulate the electron spin and the quantum interference by exploiting the atomic orbital character and the geometric profile of the nanochannel. 

Similarly to the spin configurations occurring in single-band Rashba spin-orbit systems, taking the paradigmatic example of an elliptically deformed quantum ring, we find spin and orbital textures with complex three-dimensional patterns in space exhibiting a tunable topological character with different type of spin and obital windings. Remarkably, for the low-energy electronic states we demonstrate that the spin and orbital windings are equal. We also show that the geometric phase can be controlled not only by the spin-orbit coupling but also by the shape of the nanostructure through a series of topological-like transitions that are strongly entangled to the windings of the spin-orbital textures.
Furthermore, considering that, nanostructuring methods at oxide interfaces have recently achieved a level of control that enable the design of nanoscale profiles \cite{caviglia18,bergeal,annadi,cheng,cheng1} with arbitrary shape \cite{levy1} and thickness, our findings anticipate a great potential for innovative device concepts where spin and orbital degrees of freedom can be employed to guide the electron transport by suitable selected system geometry. 

The paper is organized as follows. In the Sect. I we derive the effective model in the continuum for describing the propagation of electrons in ballistic one-dimensional curved nanostructure with $d$-orbital symmetry, spin-orbit coupling and orbital Rashba interaction. In Sect. II we present how the spatial propagation in the nanochannel leads to non-trivial spin-orbital textures and the resulting quantum phases acquired in a cycle. Then, we determine the conductance for different amplitudes of the orbital Rashba coupling and of the Fermi energy. 
Finally, in the Appendix we provide few useful symmetry relations for the trasmission coefficient and the demonstration that the winding numbers for the spin and the orbital angular momentum have equal amplitude when considering the propagation of electrons in a closed loop.

\section{Modelling curved nanochannel with $d-$ orbitals, atomic spin-orbit and orbital Rashba couplings}

We consider an effective microscopic model that is suitable for describing the 2DEG with broken out-of-plane inversion symmetry assuming that only $t_{2g}$ orbitals, i.e. $\{d_{xy},d_{zx},d_{yz}\}$, are active at the Fermi level ~\cite{khalsa13,fukaya1,fukaya2}. Due to the 2D confinement, the local $t_{2g}$ are split by the crystal field potential which favors the $xy$ configuration as the lowest one in the orbital hyerarchy.
When considering the electrons moving in a narrow quantum nanochannel we assume that only the states close to the $\Gamma$ point contribute to the electronic transport and thus the low momentum excitations can be effectively described by an Hamiltonian in the continuum that in the $\{d_{xy},d_{zx},d_{yz}\}$ orbital basis for each spin configuration is expressed as:
\begin{equation}\label{eq.cont}
H_{2D} = \int d^2 \mathbf{r}  \,\mathbf{c}^\dagger(\mathbf{r}) \mathcal{H}_{2D}(\mathbf{r}) \mathbf{c}(\mathbf{r})
\end{equation}
\noindent where $\mathbf c (\mathbf{r}) = \left( c_{xy,\sigma}(\mathbf{r}), c_{zx,\sigma}(\mathbf{r}),c_{yz,\sigma}(\mathbf{r}) \right)^T$ are the corresponding orbital fermionic fields. 
The electronic connectivity of the $t_{2g}$ bands is highly directional for symmetric TM-O bonds, e.g., an electron in $d_{xy}$-orbital can only hybridize with $p_x$ ($p_y$) states along $y (x)$ directions, respectively, in a square lattice geometry. Other microscopic ingredients include the atomic spin-orbit interaction and the orbital Rashba interaction that couples the momentum to the local orbital angular momentum within the $t_{2g}$ sector.
Taking into account the $t_{2g}$ hopping connectivity, the  inversion-broken mirror symmetry and the atomic spin-orbit interaction, the one particle Hamiltonian can be written as $\mathcal{H}_{2D}(\mathbf r) = \mathcal{H}_t(\mathbf r) + \mathcal{H}_{is}(\mathbf r) + \mathcal{H}_{SO}$. More specifically, the kinetic term is given by
\begin{equation}
\mathcal{H}_t(\mathbf r ) = \left(
                    \begin{array}{ccc}
                      -t_1 \nabla^2 &  &  \\
                       & -t_1 \nabla^2 &  \\
                       &  & -t_2 \nabla^2 + \Delta_t \\
                    \end{array}
                  \right)\otimes \sigma_0
\end{equation}

\noindent where $\nabla^2=\partial_x^2+ \partial_y^2$, $t_1,t_2$ are the orbital dependent hopping amplitudes, for a surface layer $\Delta_t$ denotes the crystal field potential. 
We point out that for the case of the orbital dependent motion along the curved nanochannle, the effective mass would a priori depend on the propagation direction. Here, since we focus on the low density regime where the most isotropic $xy$ band is dominating and we are interested in extracting the consequences arising in the phase coherent transport from the interplay of spin and orbital polarizations via the atomic spin-orbit and the orbital Rashba coupling, we neglect the orbital dependence of the effective mass. Then, $t_1$ and $t_2$ are assumed to be constant.

The inversion symmetry breaking term reads
\begin{equation}
\mathcal H_{is}(\mathbf r) = -i\Delta_{is} \partial_x l_y\otimes \sigma_0 + i \Delta_{is} \partial_y l_x\otimes \sigma_0
\end{equation}

\noindent and the atomic spin orbit interaction is expressed as
\begin{equation}
\mathcal H_{SO} = \lambda_{SO} \mathbf l \cdot \boldsymbol \sigma
\end{equation}

\noindent where we have introduced the matrices for the projected $l=1$ angular momentum associated with the $d$-states in the $t_{2g}$ symmetry sector 
\begin{equation*}
l_x = \left(
                    \begin{array}{ccc}
                      0 & 0 & 0 \\
                      0 & 0 & i \\
                      0 & -i & 0 \\
                    \end{array}
                  \right) \, l_y = \left(
                    \begin{array}{ccc}
                      0 & 0 & -i \\
                      0 & 0 & 0 \\
                      i & 0 & 0 \\
                    \end{array}
                  \right) \,
l_z = \left(
                    \begin{array}{ccc}
                      0 & i & 0 \\
                      -i & 0 & 0 \\
                      0 & 0 & 0 \\
                    \end{array}
                  \right)
\end{equation*}

When the split energy $\Delta_t$ of the $xy$ and the $\{zx,yz\}$ orbitals at the $\Gamma$ point is negative, at the first order in the orbital Rashba field there is a $k$-linear dependence spin splitting in the lowest band around $\Gamma$, so that the spin state can be effectively described by a Rashba model with Rashba coupling $\alpha = -\frac{\lambda_{SO}\Delta_{is}}{\Delta_t t_2} $~\cite{khalsa13}. 
We will show that for a curved nanochannel, due to the geometrically induced orbital texture, this description cannot be strictly applied for exctracting information on the quantum phase and on the electronic transmission.
Hereafter, we will employ the parameter $\alpha$ to set out the strength of the Rashba interaction.
Moeover, since we consider electrons which are spatially confined to move along a curved narrow channel, it is expected that the propagation of the spin state is affected by the non-uniform spatial curvature of the wire. For instance, the transmission in a Rashba semiconductor the spin direction winds with respect to the Frenet-Serret axes due to a non-homogenous curvature, allowing to control the geometric phase produced~\cite{ying}. 

The shape of the curved nanochannel can be described by introducing two unit vectors $\hat{\mathcal T}(s)$ and $\hat{\mathcal N}(s)$, which are tangent and normal to the spatial profile at a given position labelled by the curvilinear arclength coordinate $s$.
They can be expressed in terms of a polar angle $f(s)$  as  $\hat  N(s) = (\cos f(s),\sin f(s),0 )$ and $\hat T(s) = (\sin f(s),-\cos f(s),0 )$, and are related via the Frenet-Serret type equation $\partial_s \hat N = \kappa \hat T$ with $\kappa (s) = - \partial_s f(s)$ being the local curvature.

Following the Refs.~\cite{dacosta81,gentile13}, one can generally consider a local frame $(T(s),N(s))$ and introduce the curvilinear coordinates $(q^1,q^2)=(s,u)$ to re-express the kinetic operator.
The Laplacian in these curvilinear coordinate can be expressed as $\nabla^2 =\frac{1}{\sqrt{G}} \frac{\partial}{\partial q^i} \sqrt{ G} G^{-1}_{ij} \frac{\partial}{\partial q^j}$ with $G=\det G_{ij}$ and $G_{ij}$ being the metric tensor, having the following non zero-elements $G_{11}=(1+u\kappa)^2$, $G_{22}=1$.  The solution of the one-particle time-independent Schr\"{o}dinger equation can be written as $ \ket{\Psi(x,y)} = G^{-\frac{1}{4}} \ket{\Psi'(s,u)}$ so that $\int \braket{\Psi'(s,u)}{\Psi'(s,u)} dsdu=1$, and for a curved narrow wire one is taking the limit $u\to 0$. 

Hence, we have that $\nabla^2 \ket{\Psi} \sim (\partial_s^2 + \partial_u^2 + \frac{\kappa^2}{4})\ket{\Psi'}$, and in our analysis we will neglect the geometric potential coming from the term $\kappa^2/4$. The broken-inversion term gives $\mathcal H_{is}\ket{\Psi} \sim -i\Delta_{is} \left(l_N \partial_s+l_T\left(\frac{\kappa}{2}-\partial_u\right) \right)\ket{\Psi'}$. We consider a confinement potential $V(u)$ even in $u$. 
The Schr\"{o}dinger  equation can be solved by employing an adiabatic approximation, so that $\ket{\Psi'} \sim  \Phi(u) \ket{\psi(s)}$ with $\Phi(u)$ solution of $\left( -t_2 \partial_u^2+ V(u)\right) \Phi(u) = E_{\Phi} \Phi(u)$~\cite{note1}. The wave function $\ket{\psi}$ is solution of $\mathcal{H} \ket{\psi(s)} = E \ket{\psi(s)}$ where the effective Hamiltonian is obtained by averaging with respect to the ground state $\Phi(u)$ and reads~\cite{note2}

\begin{eqnarray}
\nonumber \mathcal{H}&=& -\left(t_2\frac{\mathbf{l}^2}{2}+\left(t_1-t_2\right)l_z^2 \right)\partial_s^2 + \Delta_t\left(\frac{{\mathbf{l}}^2}{2}-l_z^2 \right) \\
 && - i \frac{\Delta_{is}}{2} (l_N \partial_s + \partial_s l_N)+\lambda_{SO}\mathbf{l}\cdot\boldsymbol{\sigma} \label{eq.hamiltonian}
\end{eqnarray}

We remark that the Hamiltonian is symmetric with respect time-reversal transformation, i.e. $\Theta \mathcal{H} \Theta^{-1} = \mathcal{H}$ where $\Theta$ is the anti-unitary operator $\Theta = i \sigma_y \mathbf K$ and $ \mathbf K$ is the operation of complex conjugation.

We observe that for a circular wire, the time-independent Schr\"{o}dinger equation $\mathcal{H}\ket{\psi_{n\sigma}} = E_{n\sigma} \ket{\psi_{n\sigma}}$ can be solved by taking in account the rotational symmetry with respect to the $z$ axis. In this case the normal direction is radial $\hat{\mathcal N}=\hat R$, and the polar angle is $f(s)=s/R$ where $R$ is the radius $R=L/2\pi$. The eigenfunction $\ket{\psi_{n\sigma}}$ with energy $E_{n\sigma}$ can be then expressed as
\begin{equation}\label{eq.sol circle}
\ket{\psi_{n\sigma} (s)} = e^{i \left(n+\frac{1}{2}\right)\frac{s}{R}} U_z\left(s/R\right) \ket{\psi_{n\sigma}(0)}
\end{equation}

\noindent where $U_z(\theta)= e^{i (l_z - \frac{\sigma_z}{2})\theta}$ and $n$ is  an integer for periodic boundary conditions.

In particular the state $\ket{\psi_{n\sigma}(0)}$ and the eigen-energy $E_{n\sigma}$ are given by the eigenvalues equation

\begin{eqnarray*}
&& \bigg\{ \frac{1}{R^2}\left(t_2\frac{\mathbf{l}^2}{2}+\left(t_1-t_2\right)l_z^2 \right)\left(l_z-\frac{\sigma_z}{2}+n +\frac{1}{2}\right)^2 \\
&& +\Delta_t\left(\frac{{\mathbf{l}}^2}{2}-l_z^2 \right) + \frac{\Delta_{is}}{R}\left[ l_x\left(l_z-\frac{\sigma_z}{2}+n +\frac{1}{2}\right)-i \frac{l_y}{2}\right] \\
&& + \lambda_{SO} \mathbf{l}\cdot\boldsymbol{\sigma}\bigg\} \ket{\psi_{n\sigma}(0)} = E_{n\sigma} \ket{\psi_{n\sigma}(0)}
\end{eqnarray*}

We note that the Kramers conjugate of the state $\ket{\psi_{n\sigma} (s)}$ is $\Theta \ket{\psi_{n\sigma} (s)} = \ket{\psi_{-n-1 \sigma} (s)}$, so that  $E_{-n-1 \sigma}=E_{n\sigma}$.

\section{Electron transport in curved nanochannel with $d$-orbital symmetric states}

We consider an electron with energy $E$ injected at the extremity of the narrow quantum wire which we indicate as $s=0$.
The spatial propagation for an arbitrary curvature of a nanowire can be achieved by employing a rectification procedure by taking the limit of infinitesimal segments (for instance see Ref.~\cite{bercioux05}). Here, we consider a wire of length $L$ (in unit of the inter-atomic lattice distance) and we construct the solution by solving the model Hamiltonian in each segment. Locally a solution of the time-independent Schr\"{o}dinger equation is expressed as $e^{i s k} \ket{\chi(s,k)}$ where we refer to $\ket{\chi(s,k)}$ as the polarization state at a given position $s$.

The polarization states are related through the relation $\ket{\chi(s,k)}=U_z(f(s)) \ket{\chi(k)}$, where the polarization $\ket{\chi(k)}$  at the input point $s=0$ and the value of $k$ satisfy the equation $H_0(k)\ket{\chi(k)}=E \ket{\chi(k)} $, with

\begin{eqnarray}
\nonumber H_0(k)  &=& \left(t_2\frac{\mathbf{l}^2}{2}+\left(t_1-t_2\right)l_z^2 \right)k^2 + \Delta_t\left(\frac{{\mathbf{l}}^2}{2}-l_z^2 \right) \\
\nonumber && + \Delta_{is} l_x k +\lambda_{SO}\mathbf{l}\cdot\boldsymbol{\sigma}
\end{eqnarray}

Due to the Kramers degeneracy the values of $k$ are  $\{k_\sigma,-k_\sigma\}_\sigma$, where $k_\sigma$ gives a forward propagating solution. 

We indicate with $\Pi(s)$  the projector to the subspace $\mathbb{V}(s)$ spanned by the states $\{\ket{\chi(s,k_\sigma)}\}_\sigma$. From the continuity of the wave function, the spatial propagation of the input polarization state is determined by the spatial propagator $\Gamma(s)$

\begin{equation}\label{eq.gamma}
\Gamma(s) = \mathcal{P} e^{i \int_0^s G(s') ds'} \Pi
\end{equation}

\noindent where $\Pi=\Pi(0)$ is the projector at $s=0$, $\mathcal{P}$ is the path ordering operator, and the generator $G(s)$ is the sum of two terms $G(s) = K(s) - A(s)$ with $K(s)$ being the generator of the propagation along a flat segment at a given position $s$, and can be expressed as $K(s) = U_z(f(s))  K U_z^\dagger(f(s))$ with $K=\sum_{\sigma\sigma'} k_\sigma X^{-1}_{\sigma \sigma'} \ket{\chi(k_\sigma)}\bra{\chi(k_{\sigma'})}$ where $X$ is the matrix with elements $X_{\sigma \sigma'} = \braket{\chi(k_{\sigma})}{\chi(k_{\sigma'})}$.
Conversely $ A(s)= \left[l_z-\frac{\sigma_z}{2}, \Pi(s) \right] \Pi(s) \kappa(s)$ generates the parallel transport.

We remark that a general solution of the Schr\"{o}dinger equation can be expressed as the linear combination $\ket{\psi(s)}=c_1\Gamma(s) \ket{\phi}+c_2\Theta \Gamma(s) \ket{\phi}$. Moroever, we note that the rotation of the polarization state depends on the spatial curvature. In particular for the limiting case of a circle, the curvature $\kappa(s)$ is constant, i.e. $\kappa(s) = - 2\pi/L$, and  the path ordered integral in Eq.~\eqref{eq.gamma} can be solved, so that $\Gamma\left(s \right)$ reads

\begin{equation}
\Gamma\left(s \right) = U_z(2 \pi s/L)   e^{-i\Pi(l_z-\frac{\sigma_z}{2})\Pi 2 \pi s/L +i  K s } \Pi
\end{equation}

Then, an input state $\ket{\phi}$  which is an eigenstate of $K-\Pi\left(l_z-\frac{\sigma_z}{2}\right)\Pi \frac{2 \pi}{L}$ with eigenvalue $\xi$, evolves in space as $ e^{i \xi s} U_z(2 \pi s/L)\ket{\phi} $. In particular by requiring that $\xi=(2 n-1)\pi/L$ one can obtain again the solution given in Eq.~\eqref{eq.sol circle}.
For this limiting case, the  spin $\langle \boldsymbol \sigma(s) \rangle$ and orbital $\langle \mathbf l (s) \rangle$ local orientations do not develop a tangential component. Conversely, a tangential component can emerge because of a non-homogenous curvature, allowing the orientations to wind with respect to the Frenet-Serret directions in analogy with the single orbital spin-Rashba model~\cite{ying}.

\subsection{Low-energy regime: spin-orbital textures}

We consider the electronic propagation in the simplest configuration with only two channels identified through the orthogonal polarization states $\{\ket{\chi(k_{\sigma})}\}_{\sigma=1}$. We expect that this physical configuration is paradigmatic and applies well for the case of 2DEG in the low electron filling regime when the crystal field potential is larger than the other energy scales,  $\Delta_t$ is negative and thus the $xy$ orbital is the lowest configuration.

This is indeed the case for LaAlO$_3$-SrTiO$_3$ 2DEGs or at the surface of SrTiO$_3$, whereas the typical energy scales associated to the model Hamiltonian, the microscopic parameters of our interest assumes the values $\Delta_t \sim 50-100$ meV, $\Delta_{is}\sim 20$ meV, $\lambda_{SO}\sim 10$ meV, $t\sim 200-300$ meV,\cite{khalsa13,zhong,zabaleta,salluzzo}. Furthermore, assuming quantum rings with radius larger than $100$ nm, the orbital direction $\langle\mathbf{l}\rangle$ for a given electronc configuration mainly points towards the radial direction in the low energy regime.

\begin{figure}
[!ht]
\includegraphics[width=0.99\columnwidth]{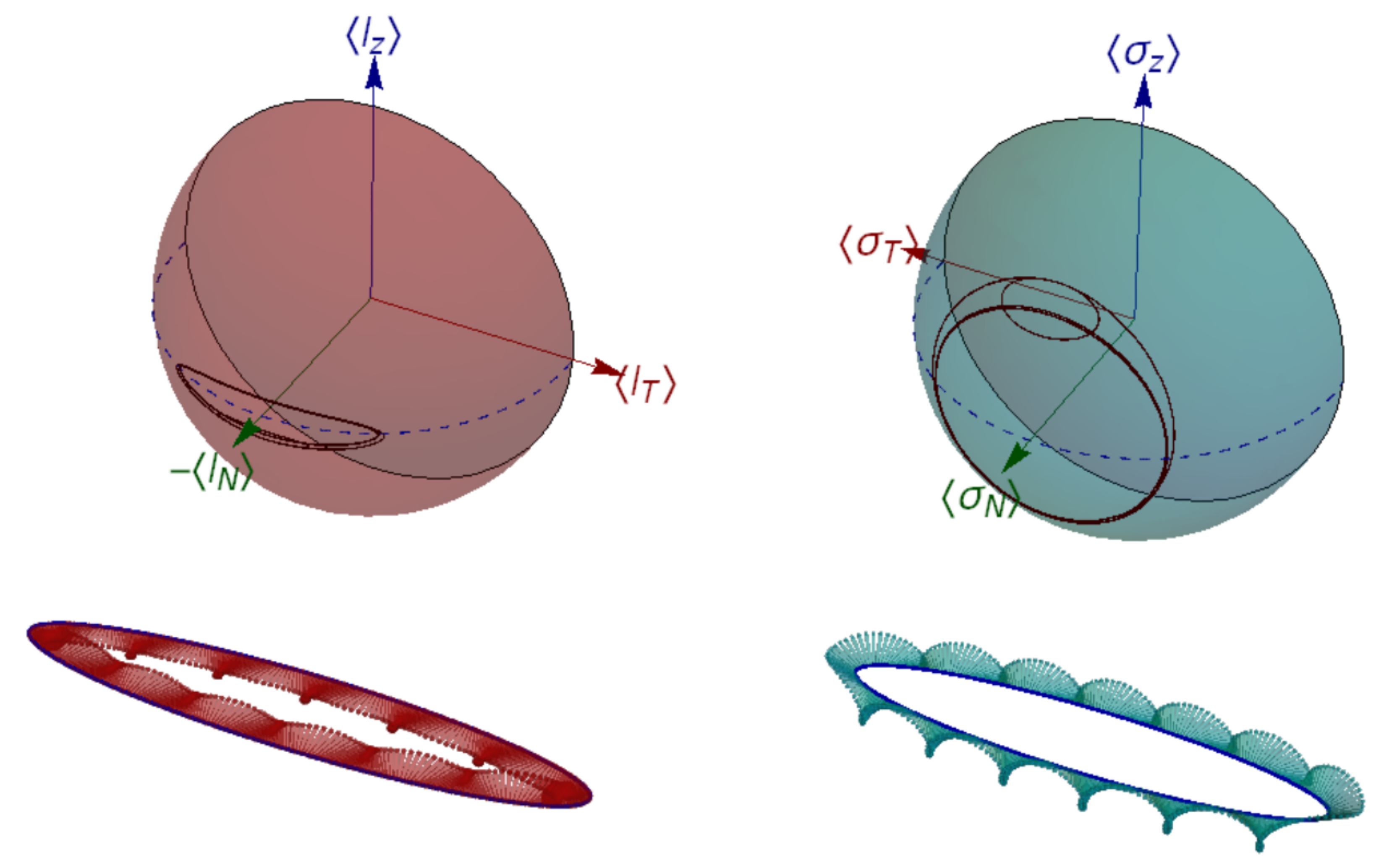}
\caption{The Bloch-Frenet-Serret (BFS) representation of $\langle \mathbf l(s) \rangle $ and $\langle \boldsymbol \sigma (s) \rangle $ for a given Fermi energy $E=-0.5315$. We put $t_1=1.$, $\lambda_{SO}=0.1$, $\Delta_{t}=-.5$, $\Delta_{is}=0.2$ in unit of $t_2$, and we consider an ellipse having a ratio of the semi-axis lengths $a/b=0.2$, with the overall length of the ring being $L= 400 \pi$ (in unit of the atomic distance).
}
\label{fig:spheres}
\end{figure}

In order to investigate the effect of the nonuniform curvature we consider the example of a quantum ring with an elliptical shape and a ratio $a/b$ between the minor ($a$) and the major ($b$) axes of the ellipse.
The propagation in space of the  spin $\langle \boldsymbol \sigma(s) \rangle$ and orbital $\langle \mathbf l (s) \rangle$ orientations is represented through two Frenet-Serret-Bloch (FSB) spheres (Fig.~\ref{fig:spheres}).

We topologically characterize the resulting spin and orbital texture through the number of windings around the normal $N$ and the binormal $z$ directions that the correspondent orientations  trace over the FSB sphere in a single loop.
Due to the reflection symmetry of the wire, these winding numbers are even, and those calculated with respect to the tangential direction are zero.
Furthermore, the spin and the orbital textures have the same winding numbers, since the relative curves are homotopic (see the Appendix for the detailed demonstration).

\begin{figure}
[!ht]
\includegraphics[width=0.99\columnwidth]{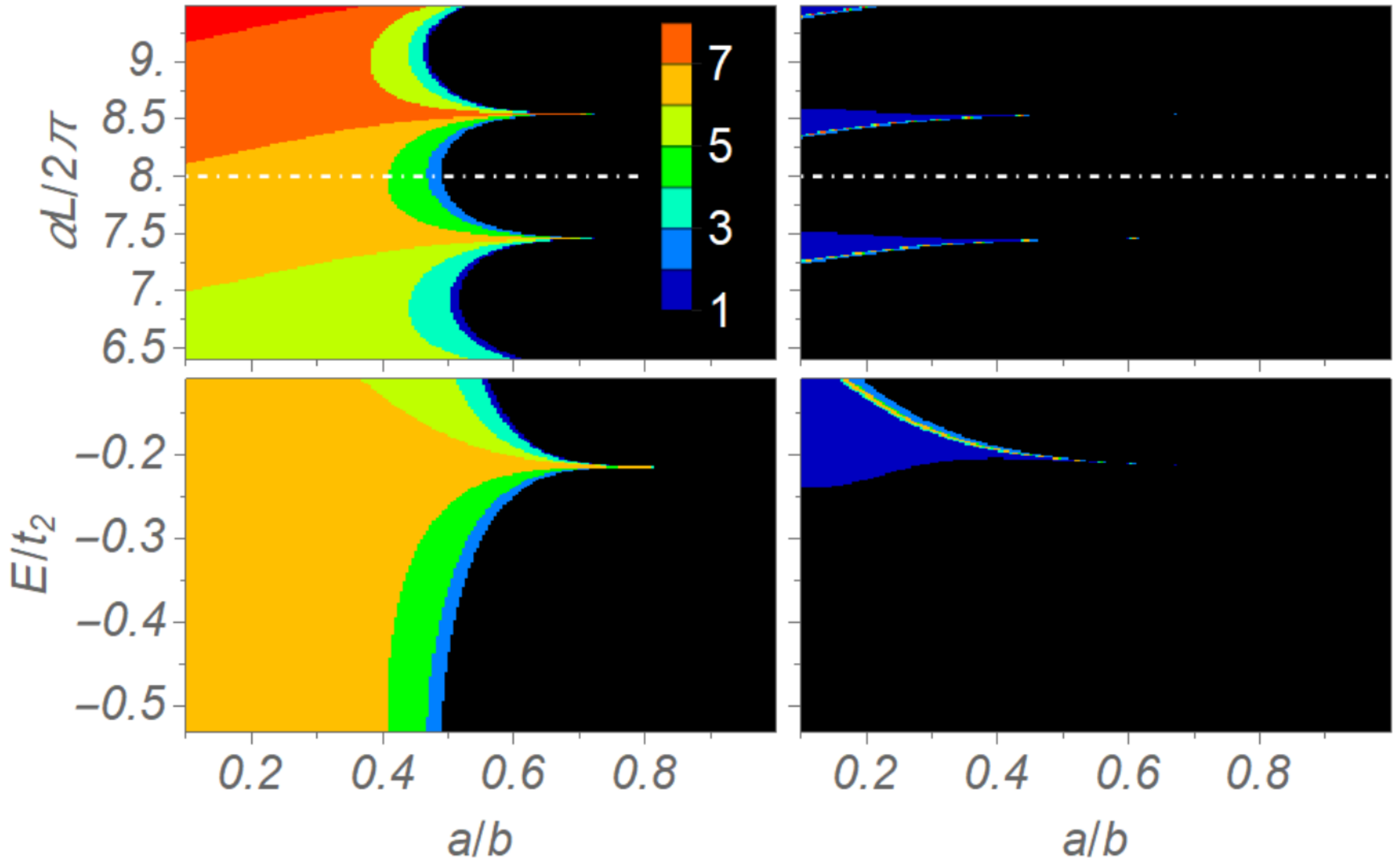}
\caption{We display the half of winding numbers calculated with respect to the normal $\mathcal N$ (left panel) and the $z$ (right panel) axes as a function of $\Delta_{is}$, the Fermi energy $E$ and the ratio $a/b$. We use the parameter $\alpha=-\frac{\lambda_{SO}\Delta_{is}}{\Delta_t t_2}$ rescaled to the length $L$ to have an adimensional amplitude that measure the effective strength of the orbital Rashba. The other microscopic parameters are set as in Fig.~\ref{fig:spheres} and $\Delta_{is}$ is changed in the interval $[0.16,0.24] t_2$. The energy $E$ is varied by fixing $\Delta_{is}=0.2$ (dotted white line).
}
\label{fig:win}
\end{figure}

We perform our analysis as a function of the inversion broken parameter $\Delta_{is}$, the Fermi energy $E$ and the semi-axis lengths ratio $a/b$.
As shown in Fig.~\ref{fig:win} the textures show two different regimes.

When the curvature is almost homogenous (black regions) the texture does not exhibit any winding in the Bloch spheres. The behavior in terms of the effective Rashba strength is similar to that one obtained for the single band Rashba model~\cite{ying}. More specifically, for a fixed energy $E$ of the selected eigenstate the winding number around the normal $\mathcal N$ increases by increasing the field $\Delta_{is}$ and the winding around the out-of-plane direction, $z$, displays a comb-like structure. Remarkably, by increasing the energy, the same patterns appear with a decreasing of the winding number around $N$.
The dependence of the spin-orbital texture on the Fermi energy is a relevant outcome of the analysis. Indeed, for the single band Rashba model the spin-orbital textures were independent of the Fermi energy. 
Here, we find that approaching a regime of strong curvature (i.e. $a/b \sim 0.4$) there are transitions in the winding which are induced by a variation of the Fermi energy.

\subsection{Low-energy regime: quantum conductance}

For a unitary propagation we can isolate the $U(1)$ contribution $e^{i \Phi}$ by writing $\Gamma(s) = e^{i \Phi} \Gamma'(s)$.
An electron moving along a closed loop can acquire a phase $\beta=\arg \bra{\phi} \Gamma'(L)\ket{\phi}$,  which can be separated as a sum of dynamical and geometric phases. The nonadiabatic Aharonov-Anandan geometric phase $\gamma_\sigma$ associated to the curve  $\ket{\phi(s)} = \Gamma'(s)\ket{\phi}$ is defined through $\gamma= \beta + i \oint \braket{\phi}{\partial_s \phi} ds $.

For a circle the  geometric phase can be expressed as the sum $\gamma = \Omega^{(s)}/2 - \Omega^{(l)}$ where  $\Omega^{(s)}$ and $\Omega^{(l)}$ are the solid angles swiped on the Bloch spheres by the spin and orbital orientations.
The phase $\beta$ enters in the transport features by considering a mirror symmetric ring with respect to the $xz$ plane, coupled to two contact leads at $s=0,L/2$.

In the limit of low bias applied voltage, the differential conductance at the energy $E$  can be obtained by means of the Landauer approach, and reads $g = e^2/h T$, where the transmission coefficient $T$ can be calculated as in~\cite{ying}. At $s=0$ the polarization states $\ket{\chi(k_{\sigma})}$ propagate along the upper arm in accordance with $\Gamma_u(s,0)=\Gamma(s)$, conversely the states $\ket{\chi(-k_{\sigma})} = \Theta \ket{\chi(k_{\sigma})}$ along the lower arm in accordance with $\Gamma_l(s,0)$ which is obtained through the time reversal $\Gamma_l(s,0)=\Theta \Gamma_u(L-s,L) \Theta^{-1}$.
We neglect backscattering effects at the contacts, and we assume that the electrons enter with same probability in lower and upper paths.

For unpolarized leads the transmission coefficient can be expressed as $T = 1 + \mathrm{Re}\Tr{\Gamma_{lu}}/M$, where we have defined $\Gamma_{lu}=\Gamma_l^\dagger\left(L/2,0\right) \Gamma_u\left(L/2,0\right)= \tilde{\Pi}\mathcal{P} e^{i \int_0^L \hat G(s')  ds'}\Pi$, where

\begin{equation}
\hat G(s) = \bigg\{   \begin{array}{cc}
                         G(s), & 0<s<L/2 \\
                         -\Theta G^\dagger(s)\Theta^{-1}, & L/2<s<L
                       \end{array}
\end{equation}

\noindent and $\tilde{\Pi} = \Theta \Pi \Theta^{-1}$. 

We note that the geometric phase is not easily related to the conductance due to the orbital contribution, indeed the counterclockwise propagation $\Gamma'(L)$ enters in the transmission through $\Gamma_{lu}\neq \Gamma'(L)$.
Indeed we observe that $\Gamma_{lu} = \Gamma'(L)$ if $\Theta {K'}^\dagger\Theta^{-1} = -K'$ and $\tilde{\Pi} = \Pi$, which is obtained if there is only one orbital $\ket{l}$. In this case $K'$ can be written as $K'= \ket{l}\bra{l}\otimes \mathbf{h}\cdot \boldsymbol \sigma$ for some $\mathbf h$ and $\Pi = \ket{l}\bra{l}\otimes \sigma_0$. 
In general the two paths are related through the relation $\Gamma_{lu} = B \Gamma'(L)$, so that the transmission coefficient $T$ can be re-expressed as

\begin{equation}\label{eq.trans}
T=1 + \textrm{Re}\sum_\sigma \bra{\phi_\sigma}B \ket{\phi_\sigma}e^{i \beta_\sigma}/M
\end{equation}

\noindent where $\{\ket{\phi_\sigma}\}_\sigma$ are the eigenstates of $\Gamma'(L)$, and those whit non zero eigenvalues are such that $\Gamma'(L) \ket{\phi_\sigma} = e^{i\beta_\sigma} \ket{\phi_\sigma}$.

At low energy only two channels are employed and the transmission coefficient $T$ can be expressed as

\begin{equation}\label{eq.trans2}
T=1 + \abs{b}\cos(\beta+\arg b)
\end{equation}

\noindent where $b = \bra{\phi_1} B \ket{\phi_1} $  which comes out due to the broken inversion, is intimately related to the orbital degrees of freedom, and can give a damping of the interference effect related to the phase $\beta$.

\begin{figure}
[!ht]
\includegraphics[width=0.99\columnwidth]{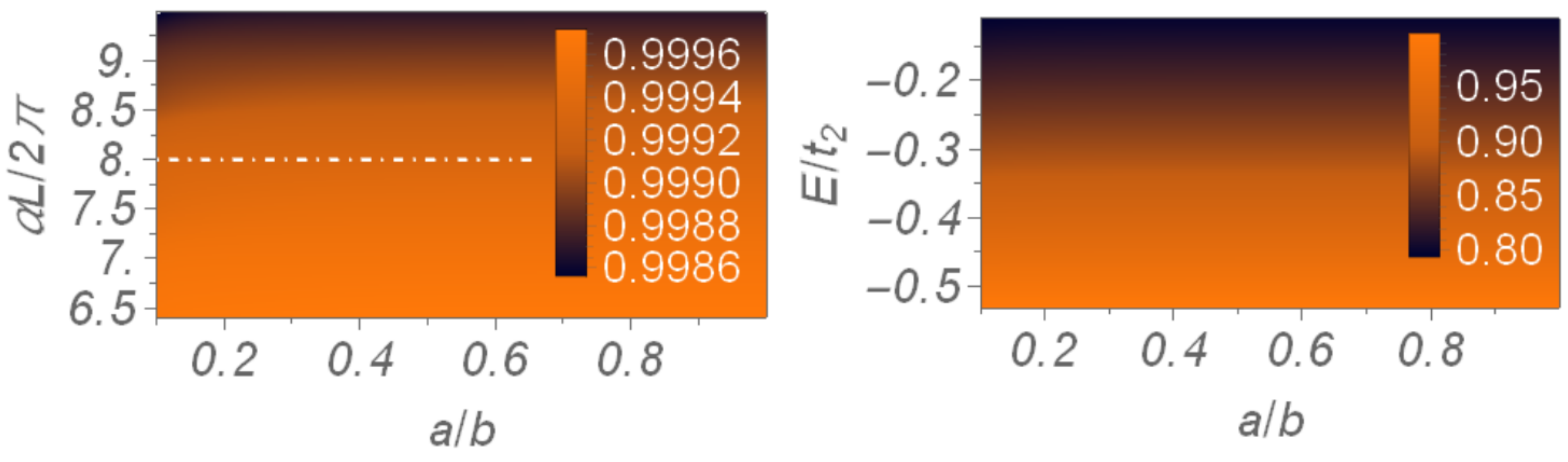}
\caption{The factor $b$ as a function of the orbital Rashba strength $\Delta_{is}$, the Fermi energy $E$ and the ratio $a/b$. The values of the other microscopic parameters are set as in Fig.~\ref{fig:win}.}
\label{fig:b}
\end{figure}

From Fig.~\ref{fig:b}, we deduce that $b$ is approximatively equal to one and sensibly decreases by increasing the energy $E$.

\begin{figure}
[!ht]
\includegraphics[width=0.99\columnwidth]{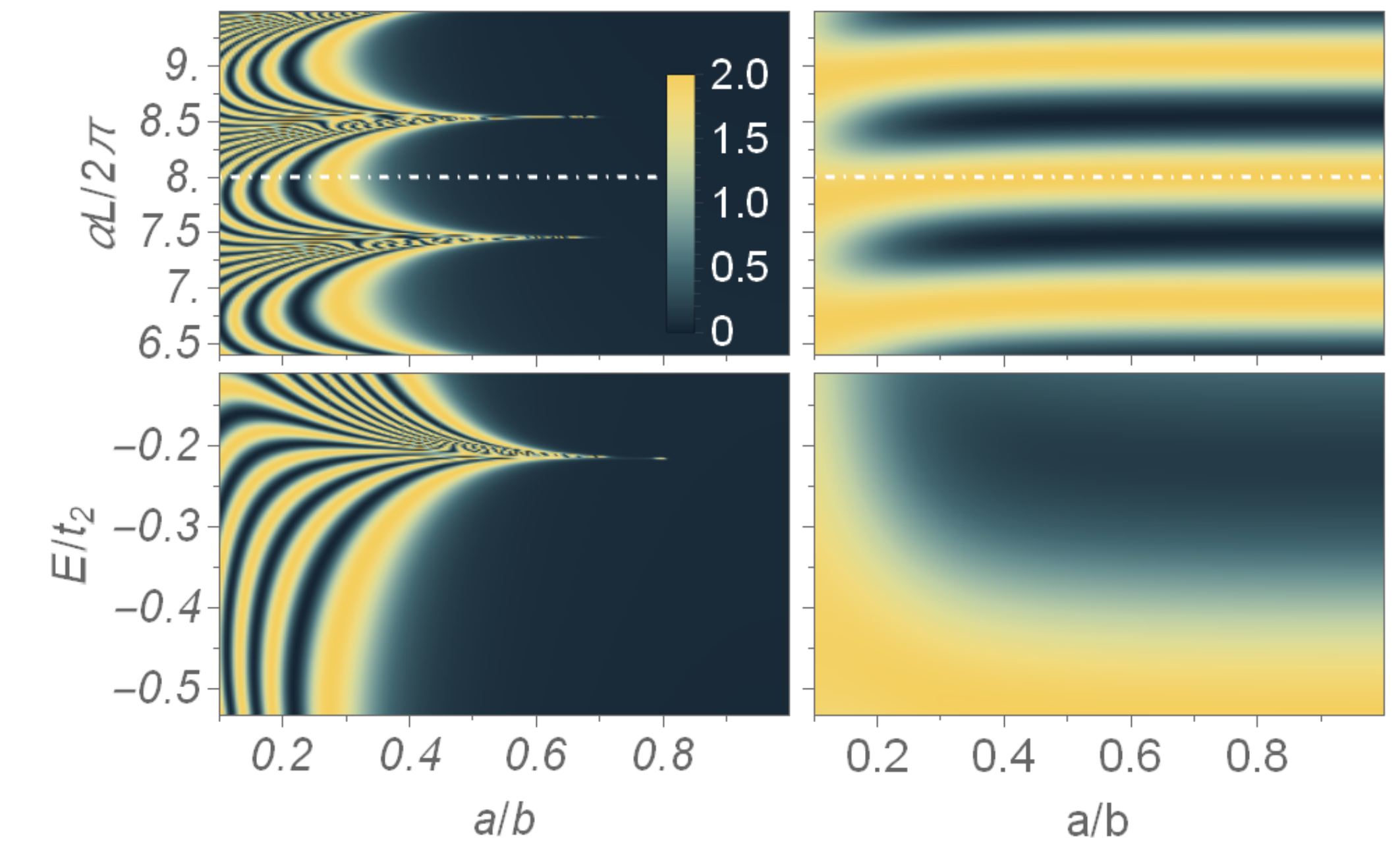}
\caption{The quantum geometric contribution $1+\cos\gamma$ (left panel), and  the transmission coefficient $T$ (right panel) as a function of the orbital Rashba coupling $\Delta_{is}$, the Fermi energy $E$ and the elliptical ratio $a/b$. The values of the microscopic parameters are set as in Fig.~\ref{fig:win}. We notice that approaching the regime of large geometrical curvature (i.e. $a/b \sim 0.4$) the evolution becomes non adiabatic and there are series of transitions with geometric phase slips. Interestingly, the total conductance can be tuned by varying the Fermi energy and thus the filling too (bottom right panel).
}
\label{fig:densi}
\end{figure}

The phase $\gamma$, the transmission coefficient $T$ in Fig.~\ref{fig:densi}.

\begin{figure}
[!ht]
\includegraphics[width=0.98\columnwidth]{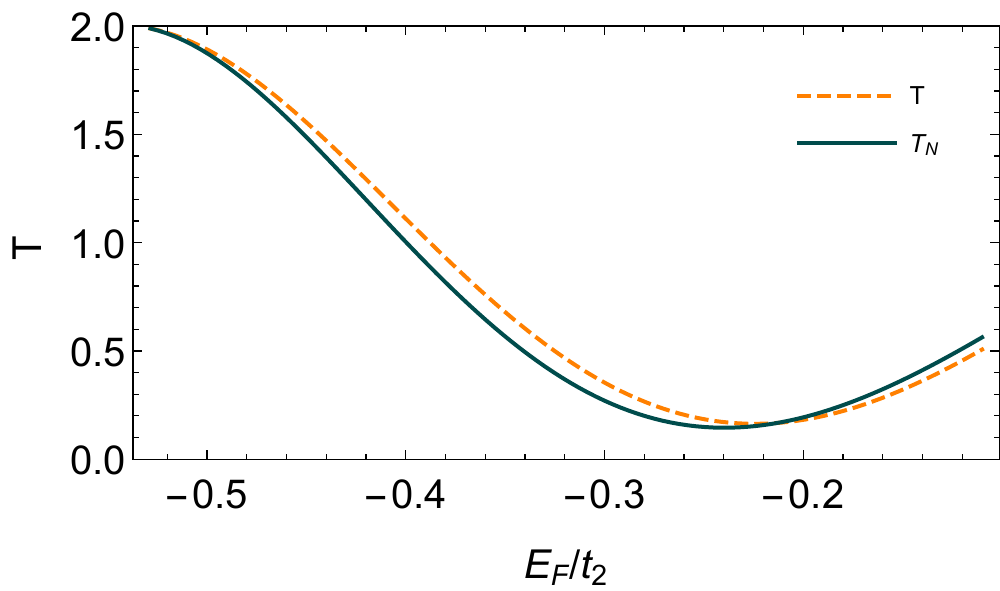}
\caption{Transmission coefficient as a function of the Fermi energy $E_F$. The values of the parameters are set as in Fig.~\ref{fig:spheres}. The trasmission $T$ is calculated through the effective continuum model for a circle of ray $R=200$, and $T_N$ by considering the Bloch Hamiltonian $H_{2D}(\mathbf k)$ of the corresponding tight binding model and by constraining the quasi-momentum $\mathbf k$ to change along the direction of the side for a circumscribed polygon of $N=100$ sides.
}
\label{fig:tra_N}
\end{figure}

As the curvature becomes nonhomogenous for small $a/b$ the geometric phase $\gamma$ sensibly differs from the adiabatic value $\pi$ in relation with the nontrivial windings around the Frenet-Serret directions, in analogy with the Rashba model~\cite{ying}. 
Conversely, the transmission pattern is smoothed, showing almost constant conductance or a changeover from destructive to constructive interference by decreasing the ratio $a/b$,

In order to go outside the low $k$ regime, we consider a wire that is made by regular polygons with $N$ sides circumscribed in a circle of radius $R$. In fig.~\ref{fig:tra_N} we compare the transmission $T$ calculated with the continuum effective model with the transmission $T_N$
for a regular polygon with $N$ sides circumscribed in a circle of radius $R$. As one can notice, the behavior is quite similar in the two approaches.

\section{Conclusions}

To conclude, we have derived a continuum model for describing the propagation of electrons in ballistic one-dimensional curved nanostructure which are marked by a strong interplay of spin-orbital degrees of freedom due to local electronic states with $d-$orbital symmetry, atomic spin-orbit and orbital Rashba couplings.
The analysis has been focused on the low electron density regime where the $xy$-state is dominant as due to the 2D confinement and the crystalline distorions.
The microscopic regime is relevant for LAO-STO 2DEG oxide nanochannels.
We find that, although the $xy$ orbital is dominant, the geometric curvature can drive both spin and orbital textures by yielding three-dimensional non-collinear patterns. Remarkably, even if the total angular momentum is not a good quantum number due to the crystal field potential and the reduced symmetries, we find that the spin and orbital angular momentum manifest the same winding when the electron travels in a closed loop. This implies that the spin and orbital components can coherently contribute to the transport properties.

Concerning the electronic transport, we demonstrate that the orbital Rashba can drive a change in the conductance and more importantly that the transmission depends on the Fermi energy. This is a relevant aspect that distinguishes the orbital Rashba coupled nanochannel from the single band Rashba case where the ballistic transport and the spin-texture are independent of the Fermi energy.
According to these results, even in a single quantum well mode nanochannel, one can expect variations of the conductance when the electron density is modified, for instance by electrical gating.

\section{Appendix}

\section{Symmetry properties of the quantum geometric phase}

The Aharonov-Anandan geometric phase $\gamma_\sigma$ associated to the curve  $\ket{\phi_\sigma(s)} = \Gamma'(s)\ket{\phi_\sigma}$ is defined through

\begin{equation}
\gamma_\sigma = \beta_\sigma + i \oint \braket{\phi_\sigma}{\partial_s \phi_\sigma} ds
\end{equation}

For two channels the spatial propagator in Eq.~\eqref{eq.gamma} takes the simpler form

\begin{equation}
\Gamma(s) = e^{i\Phi \frac{s}{L}} \Gamma'(s) =  e^{i\Phi\frac{s}{L}}\mathcal{P} e^{i \int_0^s G'(s') ds'} \Pi
\end{equation}

\noindent where $\Phi= \frac{k_1+k_2}{2}L$, and

\begin{equation}
 G'(s)= \frac{k_1-k_2}{2}\tau_z(s) - \left[l_z-\frac{\sigma_z}{2}, \Pi(s) \right] \Pi(s) \kappa(s)
\end{equation}

\noindent with $\tau_z(s)=\ket{\chi(s,k_1)}\bra{\chi(s,k_1)}-\ket{\chi(s,k_2)}\bra{\chi(s,k_2)}$. Then, it follows that $\beta_{1,2}=\pm \beta$, $\gamma_{1,2}=\pm \gamma$.

The expression for the transmission coefficient $T$ in eq.~\eqref{eq.trans2} can be derived from the general one in eq.~\eqref{eq.trans} by observing that $b_1=b_2^*$ where $b_i=\bra{\phi_i} B \ket{\phi_i}$. Indeed, we note that if there is an antiunitary operator $R$ such that $R \Gamma'(L) R^{-1} = \Gamma'(L)$ then $\ket{\phi_2}=R\ket{\phi_1}$, and if $R B R^{-1}=B$ the matrix elements are related via a complex conjugation $b_1=b_2^*=b$. In particular $R$ can be realized through the time reversal $\Theta$ by considering $k_\sigma$ as odd with respect to time reversal.


We note that for wires with a spatial profile symmetric with respect to the $xz$ plane there  is a unitary operator $Y$ such that $Y^2=1$ , $Y \Pi Y = \Pi$ and $Y G'(s) Y = - G'(L-s)$.
In order to show that, we consider
$$
Y=e^{i\psi} \ket{\chi(k_1)}\bra{\chi(k_2)}+e^{-i\psi} \ket{\chi(k_2)}\bra{\chi(k_1)}+Q
$$

where $\Pi Q\Pi=0$. By choosing $\psi$ so that

$$
e^{i2\psi} = -\frac{\bra{\chi(k_1)}l_z-\frac{\sigma_z}{2}\ket{\chi(k_2)}^*}{\bra{\chi(k_1)}l_z-\frac{\sigma_z}{2}\ket{\chi(k_2)}}
$$

then exists  $Q$ so that $l_z-\frac{\sigma_z}{2}$ changes sign under the unitary transformation $Y$. Since $f(L-s)=-f(s)$ we have that $Y G'(s) Y = - G'(L-s)$.

From $Y\Gamma'(L) Y = \Gamma'^\dagger(L)$ follows that $\ket{\phi_1}= Y\ket{\phi_2}$.

We also mention that the spatial propagator $\Gamma(s)$ allows to define a curve in the Stiefel manifold defined by the set of the frames in the subspaces $\mathbb{V}(s)$ with dimension $M=\Tr{\Pi}$. The Stiefel manifold can be regarded as a fiber bundle with the Grassmannian $\mathcal{G}(6;M)$ (the set of the $M$-dimensional subspaces) as base manifold and with the set of $M\times M$ unitary matrices as fibers.
For a closed wire one then gets a Wilczek-Zee holonomy $U$ defined as $U_{mn} = \bra{m} \mathcal{P} e^{-i\int_0^LA(s) ds} \ket{n}$ with $\{\ket{n}\}$ frame in $\mathbb{V}(0)$.

\subsection{Correlation between the spin and orbital winding numbers}

In this subsection we demonstrate that the winding of the spin and orbital angular momentum have to be the same.

An operator $\Pi(s) O(s) \Pi(s)$ is represented through a $2\times 2$ matrix $O^{(V)}(s)$, so that $\langle O(s)\rangle =  \langle O^{(V)}(s)\rangle $ where the average is calculated with respect to $\ket{\phi(s)}$. 

Due to time reversal symmetry, we have that $l_\alpha^{(V)}(s)= \mathbf v_{l_\alpha}(s)\cdot \boldsymbol \tau$ and $\sigma_\alpha^{(V)}(s)= \mathbf v_{\sigma_\alpha}(s)\cdot \boldsymbol \tau$ which are considered nonsingular for every $s$, where $\boldsymbol\tau$ is the vector of the Pauli matrices. For instance, from $\Pi(s) l_N(s) \Pi(s) = U_z(f(s)) \Pi l_x \Pi U_z^\dagger(f(s))$ we have that $l_N^{(V)}(s)$ is singular iff $ \Pi l_x \Pi$ is zero, from which it can be singular only if $\langle l_N (s)\rangle = 0$ for every $s$.

From reflection symmetries with respect to $TN$ and $zT$ planes, we have that $(\mathbf v_{\sigma_T},\mathbf v_{\sigma_N},\mathbf v_{\sigma_z})$ and $(\mathbf v_{l_T},\mathbf v_{l_N},\mathbf v_{l_z})$ are two collinear triads of mutual orthogonal vectors.

For instance, from the property of symmetry reflection with respect $TN$ plane, we have that $R_z \Pi(s) R_z^{-1} = \Pi(s)$ where $R_z$ is the reflection $R_z=\left(
                                                                                                                   \begin{array}{ccc}
                                                                                                                      1 &  &  \\
                                                                                                                       & 1 &  \\
                                                                                                                       &  & -1 \\
                                                                                                                    \end{array}
                                                                                                                  \right)\otimes\sigma_x \mathbf K$, then

\begin{eqnarray*}
&& \Tr{l_T^{(V)}(s)l_z^{(V)}(s)}=\Tr{l_T(s)\Pi(s)l_z\Pi(s)}\\
&&= \Tr{R_z^{-1}R_z l_T(s)R_z^{-1}R_z \Pi(s) R_z^{-1}R_z l_z R_z^{-1}R_z \Pi(s) }\\
&&=\Tr{R_z l_T(s)R_z^{-1}R_z \Pi(s) R_z^{-1}R_z l_z R_z^{-1}R_z \Pi(s) R_z^{-1} }\\
&&=-\Tr{l_T(s)\Pi(s)l_z\Pi(s)}
\end{eqnarray*}

\noindent since $R_z l_z R_z^{-1} = -l_z$ and $R_z l_T R_z^{-1} = l_T$, from which it results that $\Tr{l_T^{(V)}(s)l_z^{(V)}(s)}=0$ which implies the orthogonality of the vectors $\mathbf v_{l_T}$ and $\mathbf v_{l_z}$.

From the collinearity property we have that
$$
(\langle \sigma_T\rangle , \langle \sigma_N\rangle,\langle \sigma_z\rangle )=(c_T\langle l_T\rangle, c_N\langle l_N\rangle  ,c_z\langle l_z\rangle)
$$
\noindent where $c_\alpha$ is defined by $\mathbf v_{\sigma_\alpha}(s)= c_\alpha(s) \mathbf v_{l_\alpha}(s)$, and $c_\alpha(s)\neq 0$ for every $s$.

The transformation

$$
\mathbf F_t (s) = ([(1-t)\abs{c_T(s)}+t] \langle l_T(s)\rangle  , [(1-t)\abs{c_z(s)}+t]\langle l_z(s)\rangle  )
$$

\noindent with $t\in[0,1]$ is a homotopy, because of $\vert \vert \mathbf F_t (s) \vert \vert_{1} \geq \delta (\abs{\langle l_T(s)\rangle } + \abs{\langle l_z(s)\rangle })>0$ for every  $t,s$, where $\delta= \text{min}\{1,\abs{c_T},\abs{c_N},\abs{c_z}\}>0$.

Then the  curves $(\langle \sigma_T(s)\rangle ,\langle \sigma_z(s)\rangle )$ and  $(\langle l_T(s)\rangle ,\langle l_z(s)\rangle )$ wind the same number of times around the origin.
The same arguments apply also for the winding around $z$.



\begin{thebibliography}{99}
\bibitem{note1} We consider $\Delta_{is}$ and $\abs{t_2-t_1}$ small with respect the wire width.
\bibitem{note2} We notice that the value of $\Delta_t$ gets shifted by $(t_1-t_2)\int\Phi^*\partial^2_u\Phi du\approx \frac{t_2-t_1}{2}\sqrt{\frac{V^{''}(0)}{2t_2}}$.
\end{thebibliography}
\end{document}